\documentclass[num-refs]{wiley-article}
\usepackage{color}
\usepackage{ulem}

\usepackage{siunitx}

\usepackage{booktabs}
\usepackage{multirow}
\usepackage{multicol}

\usepackage{makecell}

\papertype{Original Article}
\title{An Efficient Self-supervised Seismic Data Reconstruction Method Based on Self-Consistency Learning}

\author[1,2]{Mingwei Wang}
\author[1,2]{Junheng Peng}
\author[1,2]{Yingtian Liu}
\author[1,2]{Yong Li}

\affil[1]{The Key Laboratory of Earth Exploration \& Information Techniques of Ministry Education, Chengdu, Sichuan, 610059, China}
\affil[2]{School of Geophysics, Chengdu University of Technology, Chengdu, Sichuan, 610059, China}

\corraddress{Junheng Peng, The Key Laboratory of Earth Exploration \& Information Techniques of Ministry Education, Chengdu, Sichuan, 610059, China \\ Yong Li, The Key Laboratory of Earth Exploration \& Information Techniques of Ministry Education, Chengdu University of Technology, Chengdu, Sichuan, 610059, China}
\corremail{351587229@qq.com \\ liyong07@cdut.edu.cn}

\presentadd{School of Geophysics, Chengdu University of Technology, Chengdu, Sichuan, 610059, China}

\fundinginfo{No Funding}

\begin{document}

\maketitle

\begin{abstract}
Seismic exploration remains the most critical method for characterizing subsurface structures in geophysics. However, complex surface conditions often cause a non-uniform distribution of seismic receivers along survey lines, leading to irregularly acquired seismic data, which affects subsequent processing and inversion. Prior deep learning-based seismic data reconstruction methods typically rely on datasets for supervised training. While some existing methods avoid extra data, they lack effective constraints on reconstructed data, leading to unstable performance.
In this study, we propose a self-supervised self-consistency learning strategy with a lightweight network for seismic data reconstruction. Our method requires no extra datasets, and it leverages inter-component correlations in seismic data to design a loss function, optimizing a network with only 188,849 learnable parameters. Validated on two public seismic datasets, results demonstrate our approach yields high-quality reconstruction, providing significant value for large-scale and complex seismic exploration tasks.

% Please include a maximum of seven keywords
\keywords{Seismic Data Reconstruction, self-supervised Deep Learning, Seismic Data Processing, Image Processing}
\end{abstract}

\section{Introduction}

Over the past few decades, seismic exploration has been one of the most effective means of understanding subsurface structures, achieving significant success in fields such as oil and gas development \cite{1}, regional geological studies \cite{2}, and crustal research \cite{3}. Seismic exploration mainly analyzes the reflected or refracted waves from underground, and their reception requires the arrangement of receivers along the survey line on the surface. However, the conditions on the surface are always very complex, and the undulations of the terrain and the distribution of surface objects can pose significant challenges to the layout of the entire observation system. By using some proposed methods, we can eliminate the impact of terrain variations on observation data; these methods are typically referred to as static correction \cite{4}. Besides, there is usually another situation: due to various surface conditions, such as obstacles or areas with restricted access, some areas in the exploration region may not be able to accommodate the deployment of receivers, resulting in an irregular spatial arrangement of the observation system \cite{5}. The seismic data observed by receivers with irregular spatial distribution is incomplete, which can severely hinder subsequent data processing and even the entire seismic exploration work. 

To address the issue of reconstructing seismic data from irregular spatial sampling, numerous researchers have proposed various methods. Various methods based on sparse transformations are the most widely used approaches \cite{6}. The assumption is that complete regular sampling of seismic data can be sparsely represented, while irregular spatial sampling affects this sparsity, resulting in many small amplitude coefficients in the sparse domain. By filtering out the small-amplitude coefficients, the reconstruction of the seismic data can be achieved. Typically, to achieve satisfactory results, the above steps need to be repeated multiple times. For such methods, they can be classified based on the type of sparse transformation basis functions, including Fourier transform \cite{7}, curvelet transform \cite{8}, wavelet transform \cite{9}, etc, and different choices exhibit varying sparse representation performance. In addition to using fixed basis functions, many researchers have proposed using learnable basis functions as an alternative; those methods are commonly referred to as dictionary learning \cite{10} \cite{11}. However, the greatest challenge of methods based on sparse transformations lies in the filtering of small amplitude coefficients, which typically relies on manually set parameters. Achieving satisfactory results often requires multiple iterations to determine the relevant parameters. Besides the aforementioned methods, methods based on rank-reduction have also received significant attention from researchers \cite{12} \cite{13} \cite{14}. These methods assume that seismic data possesses a low-rank structure, and irregular spatial sampling disrupts this low-rank structure, leading to an increase in the rank of the data. Singular value decomposition (SVD) is the most commonly used rank-reduction operator for extracting the singular matrix of seismic data, followed by imposing constraints on the singular values. These methods also face challenges related to parameter selection. 

In recent years, deep learning \cite{15} has been widely applied in various fields, and such applications include image processing \cite{16}, speech processing \cite{17}, and text translation, where deep learning has significantly improved performance and efficiency. Deep learning methods have also been widely applied in seismic data processing, including tasks such as noise attenuation \cite{18} \cite{19} \cite{20}, fault identification \cite{21} \cite{22}, seismic reconstruction \cite{23} \cite{24} \cite{25}, and seismic inversion \cite{58} \cite{59}. In a strict sense, most work using deep learning for seismic data reconstruction is not significantly different from seismic data noise processing. It treats the missing areas of data as targets for regression, effectively transforming the entire task into a regression problem. However, inspired by some works in the field of image generation \cite{26} \cite{27}, many researchers have begun to introduce deep learning generative models into the seismic reconstruction domain \cite{28} \cite{29} \cite{30}. Essentially, various traditional computational methods reconstruct seismic data by examining the internal relationships between observed data and the spatial relationships between observed and target data, while the methods based on generative models treat the observed irregular spatially sampled data as constraints for generating seismic data in the missing areas \cite{31} \cite{32}. One important reason for the success of deep learning is the use of additional datasets for training; the large training sets provide effective constraints for the model's parameters. However, due to the lack of interpretability, the blind use of additional data, even if it yields seemingly good results, makes it difficult to trust the outcomes. At the same time, the generalization and performance of various deep learning methods largely depend on the dataset, which greatly limits the application of deep learning approaches in the reconstruction of irregularly sampled seismic data. 

Especially when processing large-scale seismic data, the shortcomings of the existing various methods will be further magnified. Various traditional computing methods usually have very low computational efficiency when dealing with large-scale data, and they often need to be processed multiple times to obtain appropriate parameter values. Deep learning methods rely on training sets, and it is difficult to construct training sets when dealing with large-scale data. One idea is to divide the data into multiple patches for processing. Although this can alleviate this problem to a certain extent, it will then face the issues of how to divide the patches and what size to choose for the patches. To solve this, we proposed a new deep learning paradigm aimed at exploring these intrinsic relationships through a simple deep learning model, called self-supervised self-consistency learning. The relationships within the data itself, which we refer to as self-consistency, are often difficult to measure directly. Besides, this self-consistency is independent of other data, so our proposed deep learning method does not require additional data for training, which is often referred to as self-supervised learning \cite{33}. Self-supervised learning does not require a large amount of additional data to construct datasets, nor does it necessitate modeling numerous data features. As a result, the number of parameters can be significantly reduced compared to traditional deep learning methods \cite{34}. Based on the proposed self-supervised learning strategy, we have designed a lightweight deep learning network, which has only 188,849 learnable parameters, enabling it to efficiently process various types of large-scale data. We applied the proposed method to the United States Geological Survey (USGS) Beaufort Sea-Arctic Alaska and the National Petroleum Reserve–Alaska \cite{48} \cite{35} seismic data to evaluate its performance. The results indicate that our proposed method can efficiently and effectively reconstruct complex and large seismic datasets, and we further observed its resistance to noise interference. This has very positive implications for large-scale seismic exploration projects, effectively addressing the data collection challenges posed by terrain limitations. 

\section{Methodology}
\subsection{Self-Consistency Learning}

Seismic exploration involves setting receivers on the surface to capture seismic signals from artificial or natural sources. However, the receiver array on the surface is not always arranged in a regular pattern, which complicates subsequent data processing. Typically, irregularly sampled seismic data can be represented as
\begin{equation}
    d = m*R,
\end{equation}
where $d$ represents the observed seismic data, $m$ is the complete seismic data that needs to be reconstructed, and $R$ denotes the spatial sampling operator of the receivers. $R$ is a matrix that contains only 0 and 1, and it is also a highly singular matrix. Therefore, the reconstruction of irregularly sampled seismic data cannot be solved directly by inverting the above equation. To address the aforementioned problem, we proposed the theory of self-consistency learning. There are varying degrees of correlation between the data received by different receivers. The reconstruction of seismic data is based on the relationship between the missing parts and the collected parts. Therefore, we proposed using a deep learning method to learn these inherent latent relationships within the data. 

We started from the classic objective function used in deep learning methods for seismic data reconstruction \cite{45} \cite{46} \cite{47}, which can be expressed as
\begin{equation}
    \theta = \arg \min_\theta \| N(d|\theta)*R-d \|^2_2
    \label{classic objective function},
\end{equation}
where $N$ represents a deep learning network and $\theta$ represents its learnable parameters. This objective function reconstructs seismic data by learning the compression and reconstruction of the collected data, and then utilizes the spatial relationship between the collected and missing data for the reconstruction. Clearly, it does not impose a direct constraint on the missing data. Thus, we addressed this problem by further constraining $N$. The constraint can be expressed as
\begin{equation}
    \theta = \arg \min_\theta (\| N(m*(1-R)|\theta)*R-d \|^2_2+\| N(d|\theta)*(1-R)-m*(1-R) \|^2_2)
    \label{improved objective function}.
\end{equation}
This objective function (equation 3) consists of two parts. We aim to achieve bidirectional prediction of irregularly sampled seismic data through a deep learning network $N$, which means that the collected data can be used to reconstruct the missing data, while the missing data can also be utilized to reconstruct the collected data. However, $m$ is the completely reconstructed seismic data we aim to obtain, so this objective function (equation 3) is self-contradictory. 

To address this issue, our research approach is to approximate and substitute $m$. Assuming that the deep learning network $N$ can effectively model the seismic data, it should satisfy the following:
\begin{equation}
    m\approx N(d|\theta),
\end{equation}
\begin{equation}
    m\approx N(m*(1-R)|\theta).
\end{equation}
By combining equations 4 and 5, we obtain
\begin{equation}
    N(d|\theta)\approx N(N(d|\theta)*(1-R)|\theta),
\end{equation}
which indicates that the reconstruction target for $N$ is consistent and the learnable parameters $\theta$ of $N$ should satisfy:
\begin{equation}
    \theta = \arg \min_\theta \|N(d|\theta)-N(N(d|\theta)*(1-R)|\theta)\|_2^2.
\end{equation}
By substituting $m$ into equation 3, we can obtain
\begin{equation}
    \theta = \arg \min_\theta (\|d-N(d|\theta)*R\|_2^2 + \|d-N(N(d|\theta)*(1-R))|\theta)*R\|_2^2 + \|N(d|\theta)-N(N(d|\theta)*(1-R)|\theta)\|_2^2).
\end{equation}
To further enhance the deep learning network's overall modeling of the data, we randomly resampled a $R^{'}$ during each iteration of the learning process to replace $(1-R)$ in the objective function. The purpose of using the aforementioned $R^{'}$ is that we hope $N$ can establish not only the relationship between the missing seismic data and the collected seismic data, but also model the internal connections between the collected data and the missing seismic data, respectively.

\begin{figure}
\centering
\noindent\includegraphics[width=\linewidth]{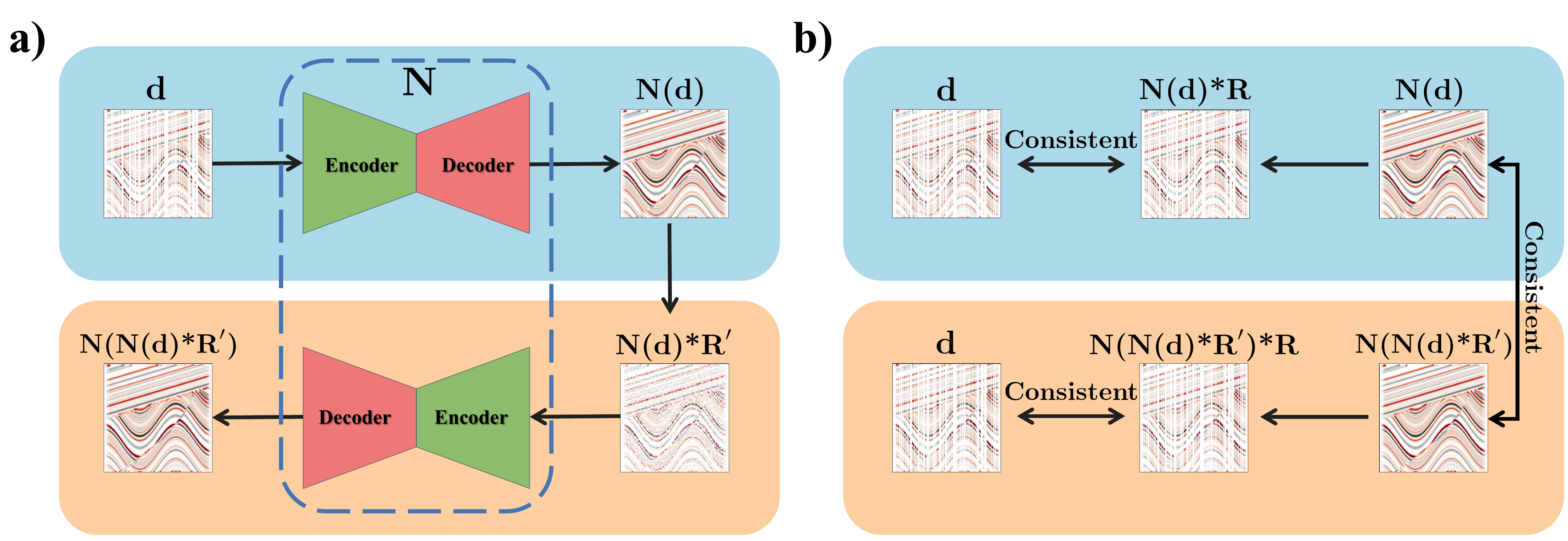}
\caption{The overall process of self-supervised self-consistency learning. a) the deep learning network $N$ is deployed twice during a single training process; b) the consistency among the input and outputs.}
\label{workflow}
\end{figure}

In summary, the self-consistency learning objective function proposed in this work can be expressed as
\begin{equation}
    \theta = \arg \min_\theta (\|d-N(d|\theta)*R\|_2^2 + \|d-N(N(d|\theta)*R^{'})|\theta)*R\|_2^2 + \|N(d|\theta)-N(N(d|\theta)*R^{'}|\theta)\|_2^2).
\end{equation}
The self-consistency loss function (equation 9) can be achieved using field observation data without the need for an additional dataset. The overall process is shown in Fig ~\ref{workflow}. 
As shown in the figure, our proposed self-supervised self-consistency learning strategy is driven solely by the internal correlations within the data itself. In Fig ~\ref{workflow}(a), the blue region represents the initially reconstructed missing data obtained through the network according to Equation 5. The orange region denotes the data that are intentionally masked again after the initial reconstruction, as described in Equation 6, and subsequently subjected to a second reconstruction. Fig ~\ref{workflow}(b) illustrates the self-consistency loss imposed between the initial reconstruction and the secondary reconstruction, as defined in Equation 7. In the next section, we provide a detailed description of the lightweight deep learning network used in our work.

\subsection{Lightweight Deep Learning Network}
\begin{figure}
\centering
\noindent\includegraphics[width=\linewidth]{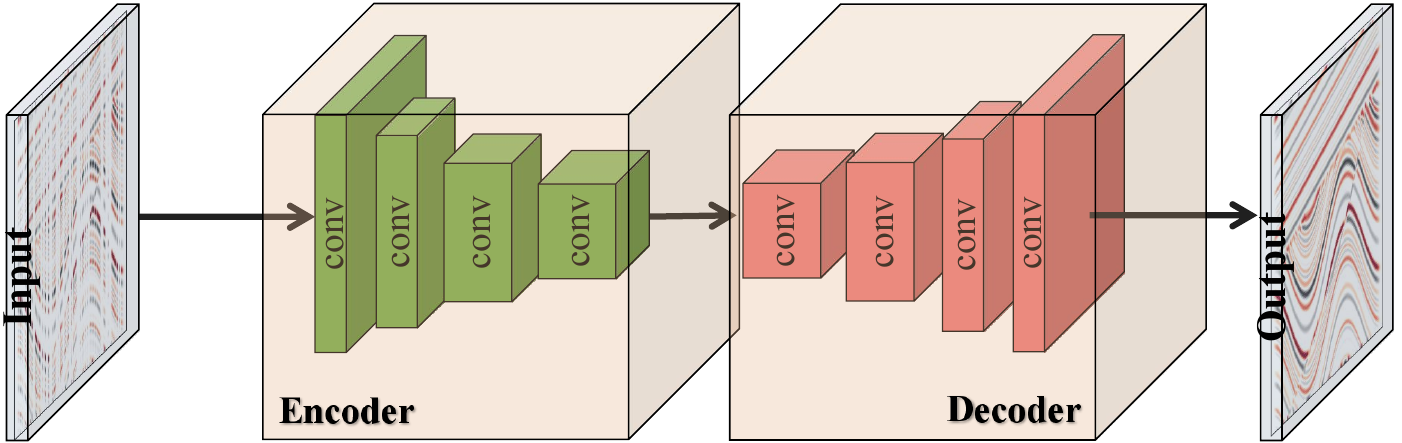}
\caption{The deep learning network we used in this work has an autoencoder structure, which includes an encoder and a decoder.}
\label{model}
\end{figure}

In this work, we adopted the structure of a convolutional autoencoder (CAE) \cite{zhang2018better}, as illustrated in Fig ~\ref{model}. The CAE has a very simple structure. First, the encoder downsamples the input seismic data while continuously expanding its channels to encode the data. The encoded data is then decoded through a fully connected layer that operates along the channel dimension. During the decoding process, transposed convolution is used to compress the channels of the data, followed by continuous upsampling to reconstruct the data. The specific parameters of each structure in the CAE are shown in Table ~\ref{hyper parameters}, and the total learnable parameters of the entire model are only 188,849, which is a decrease of several orders of magnitude compared to some other traditional deep learning networks \cite{36} \cite{37} \cite{38}. 

\begin{table}[!htbp]
\centering
\setlength{\tabcolsep}{5mm}
\renewcommand\arraystretch{1.5}
\begin{tabular}{lcccc}
\toprule[1.5pt]
\textbf{Hyper parameter} & \quad & \quad & \quad & \textbf{Setting} \\
\midrule
Encoder Channels & \quad & \quad & \quad & $[8,\ 16,\ 32,\ 64]$ \\
Kernel Size & \quad & \quad & \quad & $[4,\ 4]$, $[4,\ 4]$, $[6,\ 6]$, $[6,\ 6]$ \\
Decoder Channels & \quad & \quad & \quad & $[64,\ 32,\ 16,\ 8]$ \\
Total Learnable Parameters & \quad & \quad & \quad & 188,849 \\
\bottomrule[1.5pt]
\end{tabular}
\caption{The hyperparameters of CAE}
\label{hyper parameters}
\end{table}

\subsection{Quality Metrics}
We used multiple metrics to evaluate reconstruction quality, including Signal-to-Noise Ratio (SNR), Structural Similarity (SSIM), and R-squared ($R^2$), which are defined as follows:

\begin{equation}
SNR = 10 \cdot \log_{10} \frac{\sum_{i=1}^{N} m_i^2}{\sum_{i=1}^{N} (m_i - d_i)^2},
\label{equation: SNR}
\end{equation}
where $m_i$ and $d_i$ denote the $i$-th elements of the reference (ground truth) data and the reconstructed data, respectively, and $N$ is the total number of data points.

\begin{equation}
SSIM = \frac{(2 \mu_m \mu_d + c_1)(2 \sigma_{md} + c_2)}{(\mu_m^2 + \mu_d^2 + c_1)(\sigma_m^2 + \sigma_d^2 + c_2)},
\label{equation: SSIM}
\end{equation}
where $m$ and $d$ represent the reference (ground truth) and reconstructed data, respectively; $\mu_m$ and $\mu_d$ are the mean values of $m$ and $d$; $\sigma_m^2$ and $\sigma_d^2$ are the variances of $m$ and $d$; $\sigma_{md}$ is the covariance between $m$ and $d$; and $c_1$ and $c_2$ are small positive constants used to stabilize the division. Typically, SSIM is computed within a sliding window across the data, and the final SSIM value is obtained by averaging over all windows.

\begin{equation}
R^2 = 1 - \frac{\sum_{i=1}^{N} (m_i - d_i)^2}{\sum_{i=1}^{N} (m_i - \mu_m)^2},
\label{equation: R2}
\end{equation}
where $m_i$ and $d_i$ are the $i$-th elements of the reference and reconstructed data, respectively, and $\mu_m$ is the mean of $m$. $R^2$ measures the proportion of variance in the reference data that is explained by the reconstructed data.

Equation~\ref{equation: SNR} defines SNR, which is commonly used in signal processing to quantify the difference between the reference and reconstructed signals. Equation~\ref{equation: SSIM} defines SSIM, widely used in image processing to assess structural similarity between two datasets. Equation~\ref{equation: R2} defines $R^2$, a standard metric for evaluating the performance of regression tasks.  

SNR is highly sensitive to noise and is often used in noise attenuation and reconstruction tasks. Apart from synthetic seismic data and a small portion of field data, most field seismic datasets contain some degree of random noise, which can strongly affect reconstruction performance. In comparison, SSIM and $R^2$ are more robust to noise. Therefore, using all three metrics together provides a more comprehensive evaluation of reconstruction quality.

\section{Experiment}
In this study, we first performed random irregular spatial sampling on the complete dataset. Subsequently, we applied the proposed Self-Consistency Learning (SCL) method for data reconstruction and evaluated its performance.
For comparison, the first baseline method we adopted was a convolutional residual network based on the Swin Transformer (SCRN) \cite{gao2024swin}, for which the authors’ publicly released pretrained model was used for data processing.
To further verify the effectiveness of the proposed method, the second comparative approach employed the same CAE-based network architecture as ours but used only the mean squared error (MSE) as the self-supervised loss function (traditional self-supervised learning method), which is the traditional loss commonly adopted in conventional deep learning methods.
To assess the performance of all methods on large-scale real-world data, we conducted experiments on two publicly available large-scale seismic exploration datasets.

\subsection{Field Case 1}

\begin{figure}
\centering
\noindent\includegraphics[width=\linewidth]{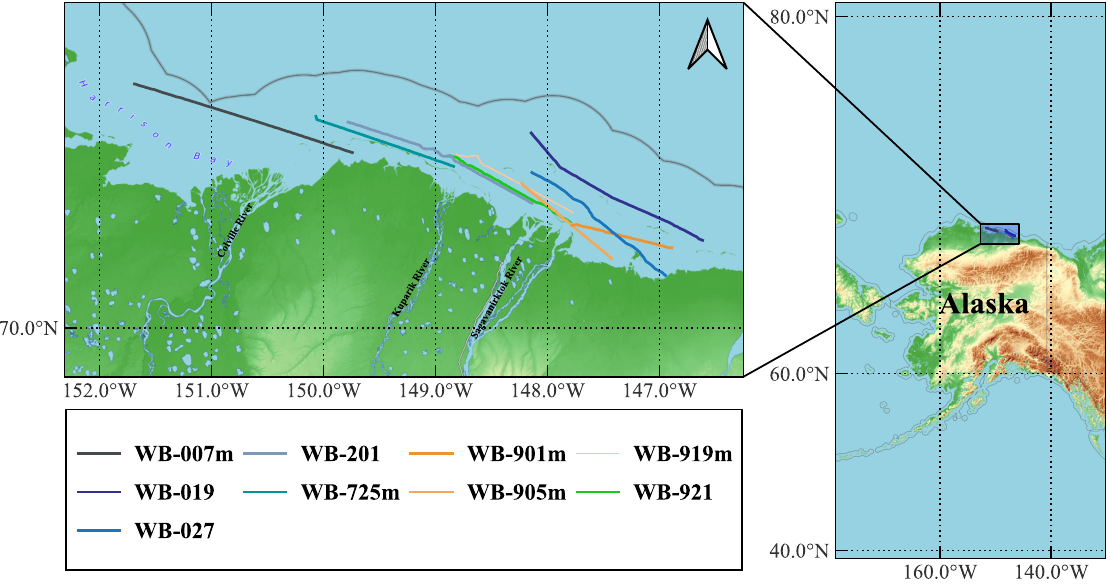}
\caption{The geographical locations of the nine survey lines of the USGS Beaufort Sea-Arctic Alaska project}
\label{location-beaufortsea}
\end{figure}

The first large-scale seismic data set we used is from the USGS Beaufort Sea-Arctic Alaska project, and it is an excellent example of data for sequence stratigraphic interpretation. It contains several geological features, including clinoforms, hinge line/zone, Barrow Arch, and bright spots. Several types of shallow faults are identified on the Beaufort shelf: high-angle, basement-involved normal faults, listric growth faults, down-to-the-north gravity faults \cite{51}. We selected nine relatively long survey lines in the entire exploration project for our evaluation experiments. Among them, the interval between each seismic trace is approximately 70 meters, and the length of single survey lines is between about 26.8 kilometers and 206 kilometers. The geographical locations of the nine survey lines are shown in Fig ~\ref{location-beaufortsea}.

We randomly removed 50$\%$ of the seismic traces among them to construct seismic data with irregular spatial sampling, and the processing results using SCL are shown in Table ~\ref{results-beaufortsea}. We compared with the traditional self-supervised learning using a classical objective function (Equation ~\ref{classic objective function}) and the SCRN method to evaluate the stable and favorable reconstruction effect of SCL. Overall, SCL has achieved better results than the traditional self-supervised learning method and SCRN method on each survey line, and the results it has obtained on different survey lines are also relatively stable. 
\begin{table}[!htbp]
\centering
\renewcommand\arraystretch{1.2}
\resizebox{\textwidth}{!}{  % 自动按页宽缩放表格
\begin{tabular}{cccccccccc}
\toprule[1.5pt]
\multirow{2}*{\textbf{Survey Line}} 
& \multicolumn{3}{c}{\textbf{SCL}} 
& \multicolumn{3}{c}{\textbf{Traditional}} 
& \multicolumn{3}{c}{\textbf{SCRN}}  \\
\cmidrule(lr){2-4} \cmidrule(lr){5-7} \cmidrule(lr){8-10}
& SNR & SSIM & $R^2$ & SNR & SSIM & $R^2$ & SNR & SSIM & $R^2$\\
\midrule
WB-007m & 11.0816 & 0.9224 & 0.9220 & 10.4397 & 0.9153 & 0.9096 & 10.9819 & 0.9200 & 0.9172 \\
WB-019  & 10.0607 & 0.9134 & 0.8991 & 9.6242 & 0.9074 & 0.8910 & 9.3118 & 0.9020 & 0.8828 \\
WB-027  & 9.6419  & 0.9098 & 0.8914 & 8.9084  & 0.9033 & 0.8714 & 9.3765  & 0.9022 & 0.8846 \\
WB-201  & 10.5066 & 0.9353 & 0.9110 & 10.4033 & 0.9342 & 0.9089 & 10.1747 & 0.9286 & 0.9039 \\
WB-725m & 10.4402 & 0.9206 & 0.9096 & 9.0981  & 0.9013 & 0.8769 & 10.0849 & 0.9137 & 0.9019 \\
WB-901m & 10.7898 & 0.9262 & 0.9166 & 7.2170  & 0.8509 & 0.8102 & 10.8519 & 0.9270 & 0.9178 \\
WB-905m & 10.4066 & 0.9195 & 0.9089 & 7.8538  & 0.8750 & 0.8361 & 10.2758 & 0.9190 & 0.9003 \\
WB-919m & 12.1215 & 0.9453 & 0.9386 & 10.0342 & 0.9155 & 0.9008 & 12.0392 & 0.9452 & 0.9321 \\
WB-921  & 12.1739 & 0.9436 & 0.9394 & 11.9464 & 0.9358 & 0.9276 & 11.7641 & 0.9393 & 0.9334 \\
\midrule
\textbf{Average} 
& \textbf{10.8025} & \textbf{0.9262} & \textbf{0.9152} 
& 9.5028 & 0.9043 & 0.8814 & 10.5401 & 0.9219 & 0.9082 \\
\bottomrule[1.5pt]
\end{tabular}
} % 结束 resizebox
\caption{The reconstruction results (with 50\% traces missing) of three methods (SCL, traditional self-supervised learning method, and SCRN) on nine survey lines in the USGS Beaufort Sea–Arctic Alaska project. The reconstruction quality is evaluated using SNR, SSIM, and $R^2$. (Bold values represent the best performance.)}
\label{results-beaufortsea}
\end{table}

\begin{figure}
\centering
\noindent\includegraphics[width=\linewidth]{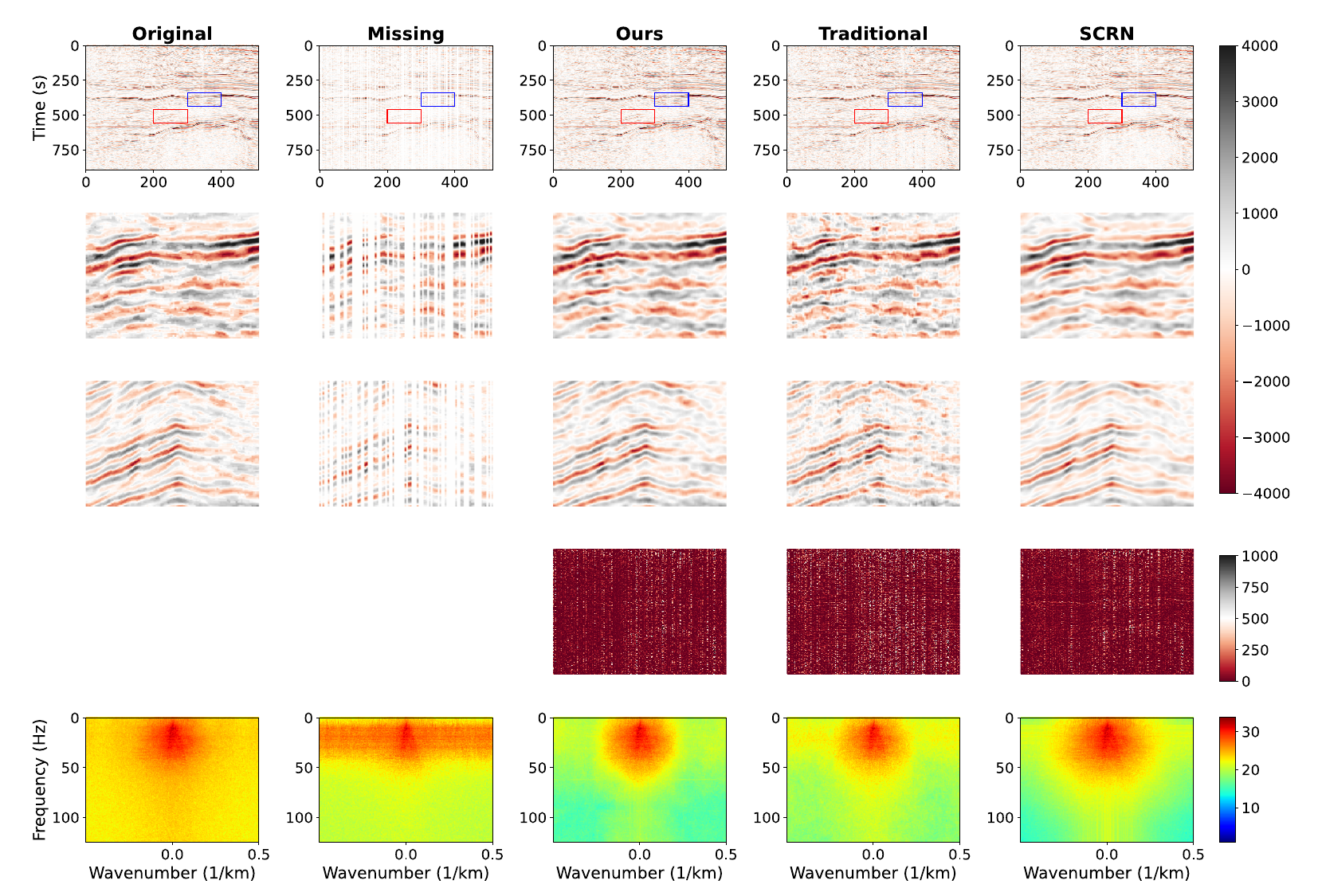}
\caption{The WB-905m survey line in the USGS Beaufort Sea-Arctic Alaska project. The first column represents the complete field collection results; the second column represents the irregular spatial sampling data used for reconstruction; the third column represents the reconstruction results of our proposed SCL; the fourth column represents the results of reconstruction using the traditional self-supervised learning method. The fifth column represents the results of reconstruction using the SCRN method. The first row presents the complete region, where the blue and red rectangles indicate two local regions enlarged in the second and third rows. The fourth row shows the residuals between the reconstructed results and the original data. The fifth row displays the corresponding F-K spectra for each case.}
\label{bs-WB905m}
\end{figure}

In addition, two survey lines, WB-905m and WB-919m, were selected for visual evaluation, and the results for WB-905m are shown in Fig.~\ref{bs-WB905m}. The second column presents the irregularly sampled data constructed by randomly removing 50\% of the seismic traces. It is evident that this irregular sampling severely reduces the lateral continuity of the data, which can significantly affect subsequent geological structure localization and interpretation. Regardless of whether the SCL algorithm, the traditional self-supervised learning method, or the SCRN method is used, the large-scale geological structures are generally well reconstructed. However, the reconstruction results obtained by the traditional self-supervised learning method and the SCRN method still show noticeable discrepancies compared with the original data. Therefore, we selected the regions marked by red and blue boxes for a detailed comparison of the reconstruction results. Compared with the SCL method, the SCRN method exhibits obvious interpolation artifacts in regions with large-scale trace gaps, indicating that the SCL method achieves more accurate seismic data recovery when facing severe trace loss. Similarly, the reconstructed data from the traditional self-supervised learning method still deviates from the original data. In contrast, the results reconstructed by the SCL method are much closer to the original data than those obtained by the SCRN and traditional self-supervised learning methods. Compared with other methods, the residual maps and the frequency–wavenumber (F–K) spectra in the figures also demonstrate that SCL achieves superior performance. The F–K spectra of SCL exhibit more concentrated energy.

\begin{figure}
\centering
\noindent\includegraphics[width=\linewidth]{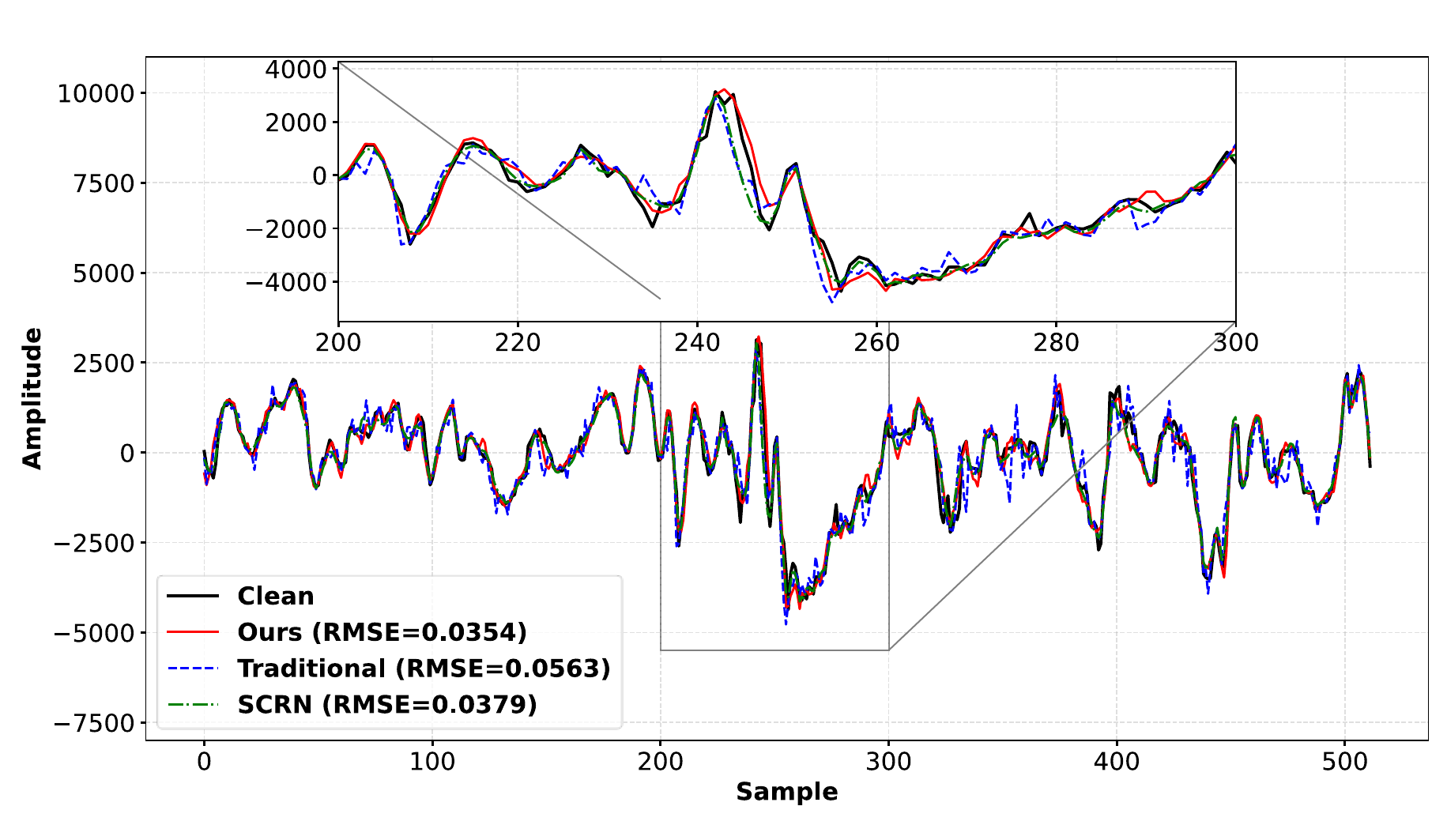}
\caption{Single-trace waveform (Trace 600) comparison of the reconstructed data by different methods for line WB-905m.}
\label{trace_wb905}
\end{figure}

To further evaluate the performance of each method, we extracted single-trace data from the reconstructed seismic sections and plotted them as waveforms, as shown in Fig.~\ref{trace_wb905}. In addition, we calculated the single-trace root mean squared error (RMSE) \cite{brassington2017mean} between the reconstructed data and the original data, which provides a more intuitive measure of the reconstruction accuracy. According to the RMSE values, the single-trace data reconstructed by the SCL method show the best agreement with the original data, followed by the SCRN method, while the traditional self-supervised learning method exhibits the largest deviation.

\begin{figure}
\centering
\noindent\includegraphics[width=\linewidth]{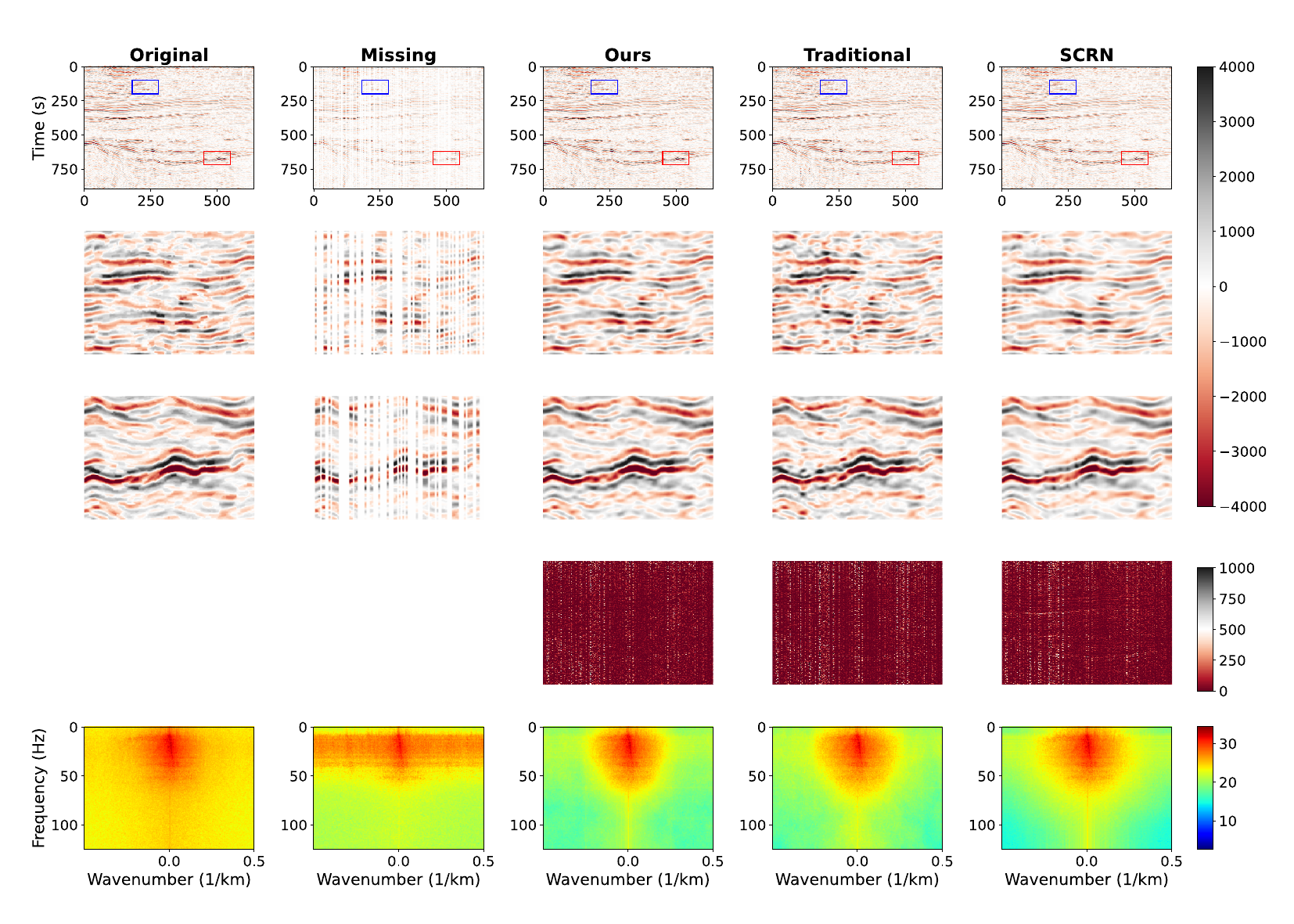}
\caption{The WB-919m survey line in the USGS Beaufort Sea-Arctic Alaska project. The first column represents the complete field collection results; the second column represents the irregular spatial sampling data used for reconstruction; the third column represents the reconstruction results of our proposed SCL; the fourth column represents the results of reconstruction using the traditional self-supervised learning method. The fifth column represents the results of reconstruction using the SCRN method. The first row presents the complete region, where the blue and red rectangles indicate two local regions enlarged in the second and third rows. The fourth row shows the residuals between the reconstructed results and the original data. The fifth row displays the corresponding F-K spectra for each case.}
\label{bs-WB919m}
\end{figure}

The same situation was also manifested in the survey line WB-919m, as shown in Fig ~\ref{bs-WB919m}. The second row in Fig ~\ref{bs-WB919m} is an enlarged display of the blue box area in the first row, where the disadvantages of the traditional self-supervised learning method are more obvious. In its processing results, the seismic event of the seismic signal has become discontinuous, and the resolution of the seismic event is even lower, which will seriously affect the interpretation work of the underground geological structure. The second row is an enlarged display of the red box area in the first row, which shows results similar to those in the second row. It can be clearly seen that the SCL we proposed can better reconstruct the seismic data, and its results are closer to the real and completely sampled seismic data compared with those of the traditional self-supervised learning method. In the third row, it can be observed that in the enlarged region of the SCRN results, obvious vertical artifacts appear when facing large areas of continuous missing traces.

\begin{figure}
\centering
\noindent\includegraphics[width=\linewidth]{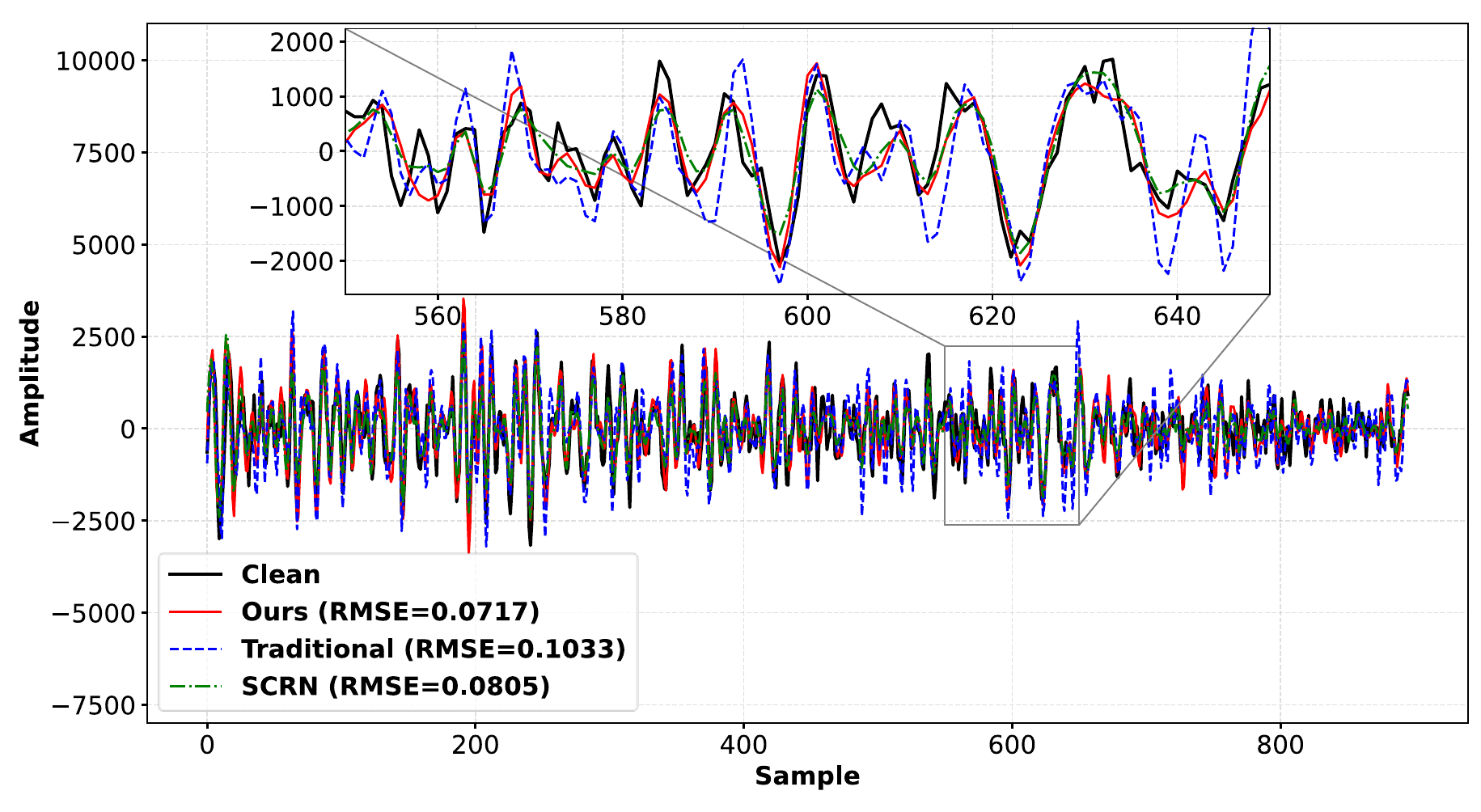}
\caption{Single-trace waveform (Trace 600) comparison of the reconstructed data by different methods for line WB-919m.}
\label{trace_wb919}
\end{figure}

Similarly, the single-trace waveform comparison for the WB-919m data is presented in Fig.~\ref{trace_wb919}. According to the RMSE evaluation, the single-trace data reconstructed by the SCL method show the closest agreement with the original data, followed by the SCRN method, while the traditional self-supervised learning method yields the largest deviation.

\begin{figure}
\centering
\noindent\includegraphics[width=0.8\linewidth]{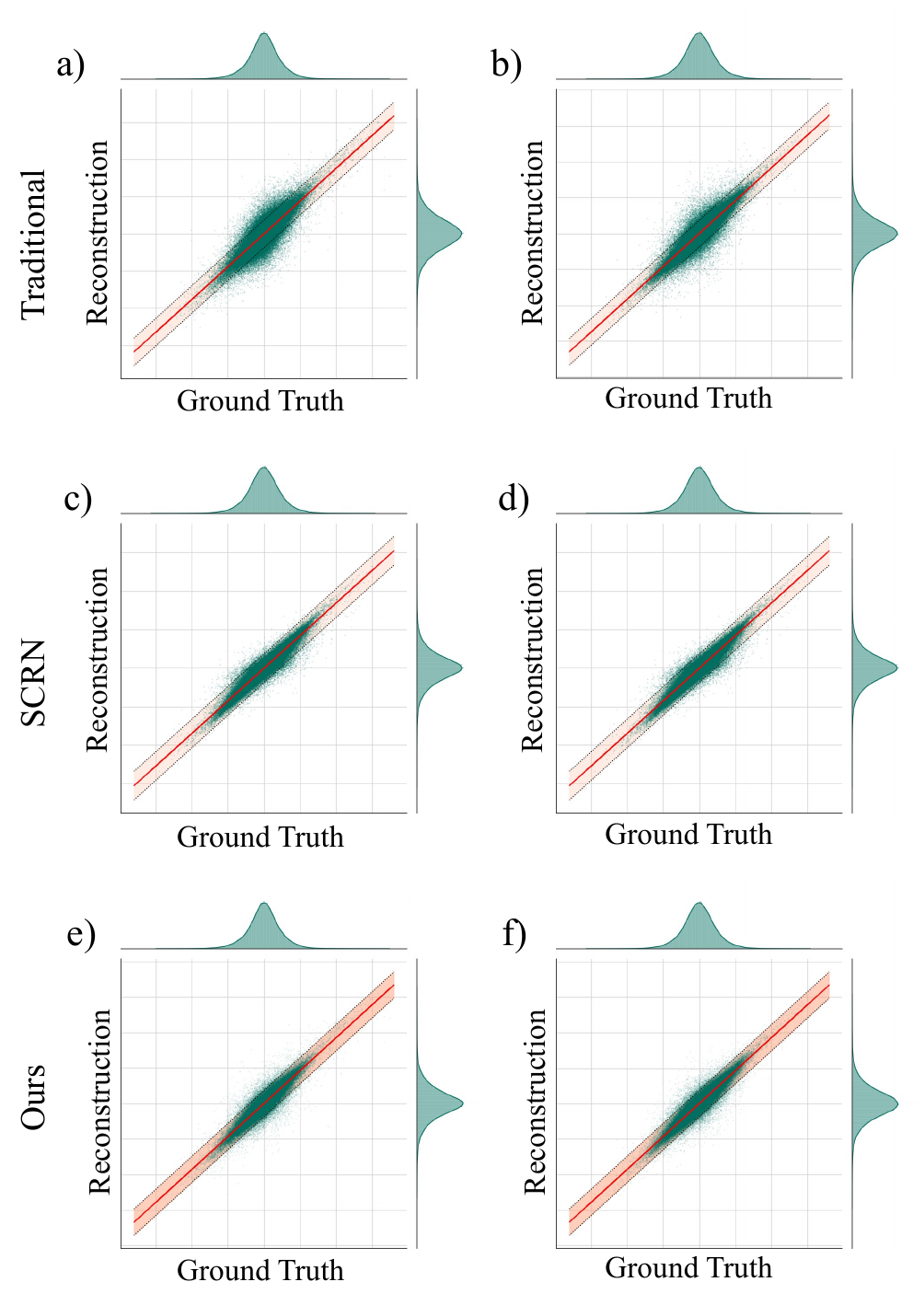}
\caption{Scatter plots of the WB-905m and WB-919m data from the USGS Beaufort Sea–Arctic Alaska project. Panels a), c), and e) illustrate the scatter relationships between the reconstructed and the original data obtained by the traditional self-supervised learning method, SCRN, and SCL for the WB-905m line, respectively. Panels b), d), and f) show the corresponding results for the WB-919m line.}
\label{bs-scatter}
\end{figure}

In addition, we present the scatter plots of two survey lines, as shown in Fig ~\ref{bs-scatter}. The scatter plots can well describe the error between the reconstructed results and the completely collected seismic data, and the closer the scatter points are to the straight line $y=x$, the closer the reconstructed results are to the completely collected data. It can be seen that, compared with the traditional self-supervised learning method, the reconstruction results of SCL are more concentrated, and the overall error is significantly smaller than that of the traditional self-supervised learning method. Due to the lack of effective constraints on the reconstructed data in the traditional self-supervised learning method, its results are less stable, and the errors are more obvious. 

\subsection{Field Case 2}

\begin{figure}
\centering
\noindent\includegraphics[width=\linewidth]{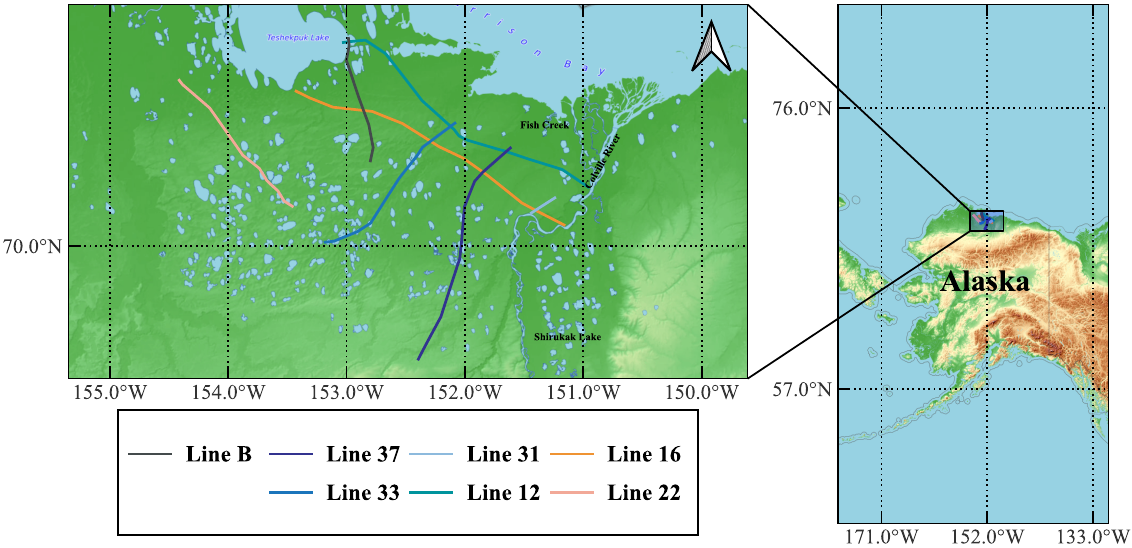}
\caption{The geographical locations of the seven survey lines of the USGS National Petroleum Reserve–Alaska project}
\label{location-centralalaska}
\end{figure}

In addition to the above-mentioned USGS Beaufort Sea-Arctic Alaska project datasets, we selected a much larger dataset from the National Petroleum Reserve–Alaska (NPRA) by the United States Geological Survey \cite{35} for further testing. The NPRA is one of the largest publicly available seismic exploration datasets at present. We selected seven representative survey lines from the original seismic dataset for the following experiments. The geographical locations of the seven selected survey lines are shown in Fig ~\ref{location-centralalaska}. Among these seven survey lines, the shortest one is Line 31, which is only about 16.6 kilometers long; the longest one is Line 16, approximately 183 kilometers long. The total length of the seven survey lines is about 815.4 kilometers; the time recording for each receiver is about 5.6 seconds, and the time recording interval is 4 milliseconds. 

\begin{figure}
\centering
\noindent\includegraphics[width=\linewidth]{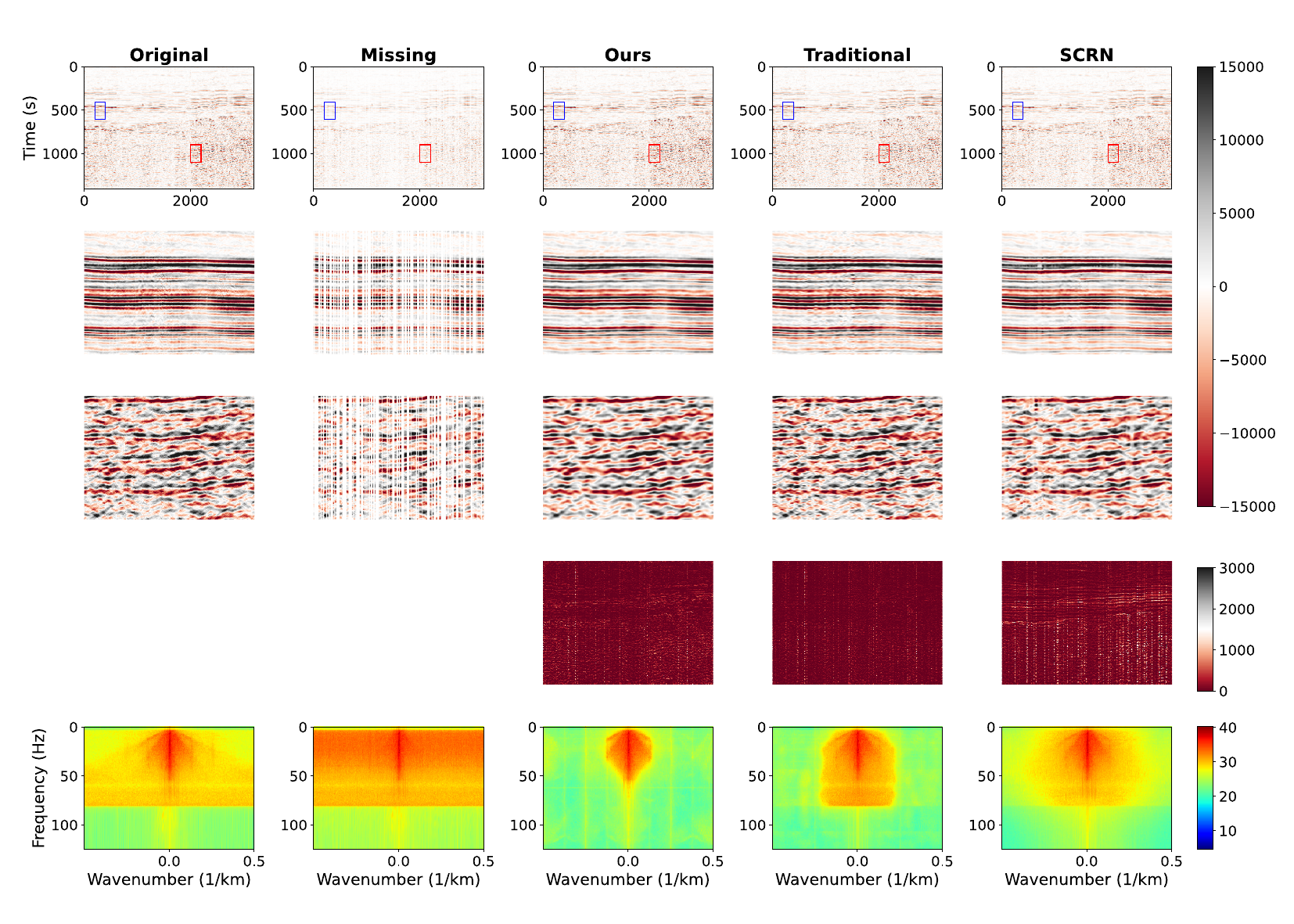}
\caption{The Line 22 survey line in the USGS National Petroleum Reserve–Alaska project. The first column represents the complete field collection results; the second column represents the irregular spatial sampling data used for reconstruction; the third column represents the reconstruction results of our proposed SCL; the fourth column represents the results of reconstruction using the traditional self-supervised learning method. The fifth column represents the results of reconstruction using the SCRN method. The first row presents the complete region, where the blue and red rectangles indicate two local regions enlarged in the second and third rows. The fourth row shows the residuals between the reconstructed results and the original data. The fifth row displays the corresponding F-K spectra for each case.}
\label{ca-Line22}
\end{figure}

Similarly, we randomly removed 50$\%$ of the seismic traces to construct irregular spatial sampling data. The processing results of the seven survey lines are shown in Table ~\ref{results-centralalaska}. Compared with the USGS Beaufort Sea-Arctic Alaska project, the noise contained in the USGS NPRA dataset is much stronger, which is unfavorable for the reconstruction work. On the seven survey lines, SCL has achieved better and more stable results than the traditional self-supervised learning and SCRN method. We select Line 22 and Line 33 for visual display to further demonstrate the reconstruction performance of SCL, as shown in Fig ~\ref{ca-Line22}. Line 22 is approximately 104 kilometers in total length. It is an onshore seismic exploration survey line, and the underground structures it contains are relatively complex. We selected two relatively typical areas among them for visual display. Among them, the second row in Fig ~\ref{ca-Line22} represents the enlarged display of the blue box area, and the third row represents the enlarged display of the red box area. There is a strong reflection interface in the blue box area, with good overall lateral continuity. However, the collected seismic data contains a certain degree of noise. The red box area is the reflection wave from a deeper structure, and the waveform of the reflection wave is more complex.
\begin{table}[!htbp]
\centering
\renewcommand\arraystretch{1.2}
\resizebox{\textwidth}{!}{  % 自动按页宽缩放表格
\begin{tabular}{cccccccccc}
\toprule[1.5pt]
\multirow{2}*{\textbf{Survey Line}} 
& \multicolumn{3}{c}{\textbf{SCL}} 
& \multicolumn{3}{c}{\textbf{traditional}} 
& \multicolumn{3}{c}{\textbf{SCRN}} \\
\cmidrule(lr){2-4} \cmidrule(lr){5-7} \cmidrule(lr){8-10}
& SNR & SSIM & $R^2$ & SNR & SSIM & $R^2$ & SNR & SSIM & $R^2$\\
\midrule
Line 12 & 11.7058 & 0.9117 & 0.9325 & 9.7270 & 0.8489 & 0.8935 & 11.3641 & 0.9011 & 0.9281 \\
Line 16 & 12.5662 & 0.9157 & 0.9446 & 11.0681 & 0.9085 & 0.9218 & 12.3131 & 0.9124 & 0.9387 \\
Line 22 & 13.1957 & 0.9247 & 0.9521 & 11.2268 & 0.8682 & 0.9246 & 12.8631 & 0.9114 & 0.9436 \\ 
Line 31 & 11.6341 & 0.9011 & 0.9314 & 9.7994 & 0.8567 & 0.8782 & 10.8710 & 0.8852 & 0.9066 \\
Line 33 & 12.5926 & 0.9172 & 0.9450 & 11.4364 & 0.9057 & 0.9282 & 11.31 & 0.9024 & 0.9194 \\
Line 37 & 13.0186 & 0.9222 & 0.9502 & 12.4882 & 0.9199 & 0.9436 & 12.8390 & 0.9211 & 0.9484 \\
Line B  & 13.5149 & 0.9212 & 0.9555 & 12.6903 & 0.9078 & 0.9462 & 12.5137 & 0.8994 & 0.9316 \\
\midrule
\textbf{Average} 
& \textbf{12.6040} & \textbf{0.9163} & \textbf{0.9445} 
& 11.2052 & 0.8847 & 0.9194 
& 12.0106 & 0.9047 & 0.9309 \\
\bottomrule[1.5pt]
\end{tabular}
} % 结束 resizebox
\caption{The reconstruction results (with 50\% traces missing) of three methods (SCL, traditional self-supervised learning method, and SCRN) on seven survey lines in the USGS NPRA project. The reconstruction quality is evaluated using SNR, SSIM, and $R^2$. (Bold values represent the best performance.)}
\label{results-centralalaska}
\end{table}

\begin{figure}
\centering
\noindent\includegraphics[width=\linewidth]{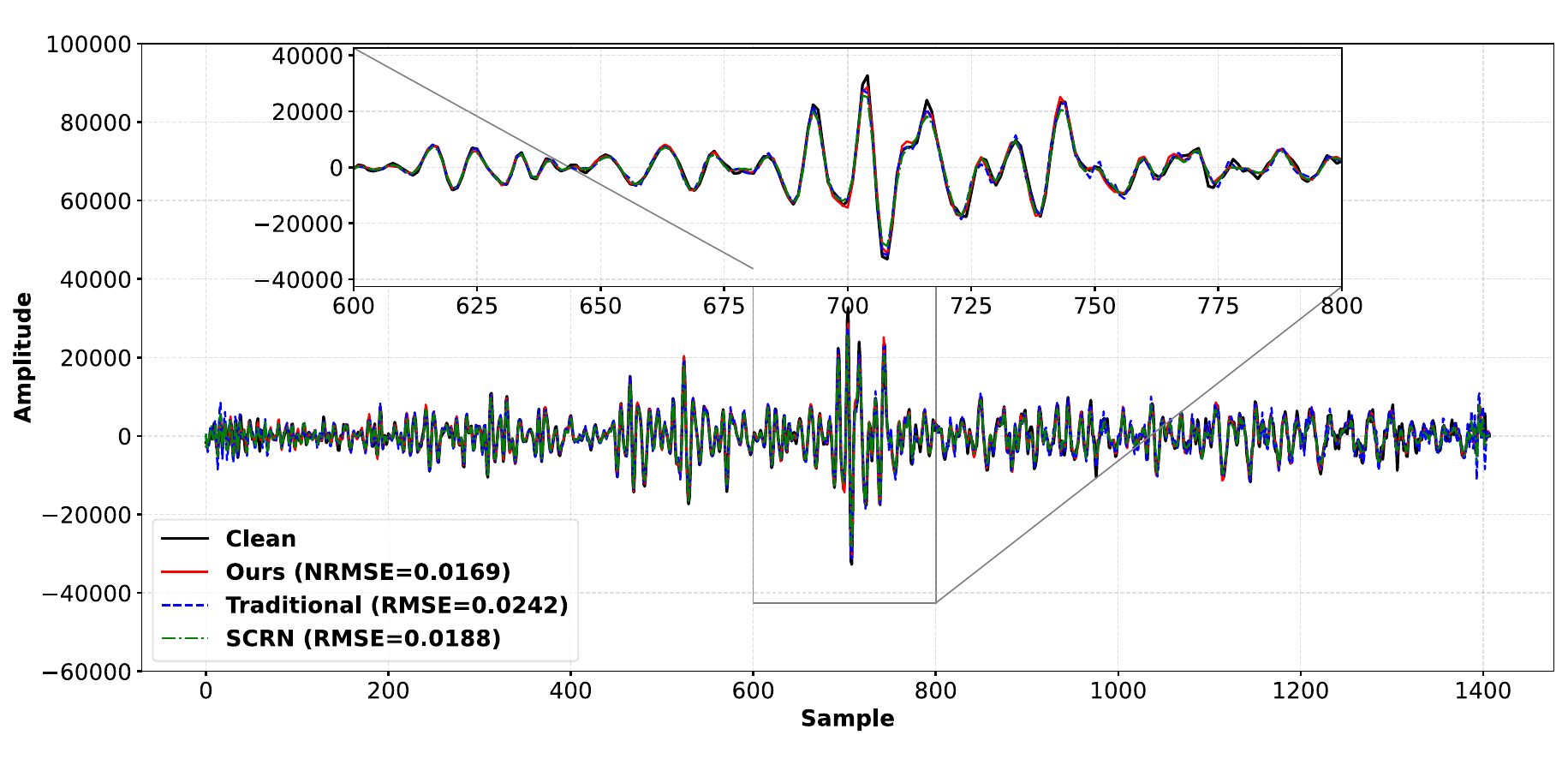}
\caption{Single-trace waveform (Trace 710) comparison of the reconstructed data by different methods for line 22.}
\label{trace_line22}
\end{figure}

In the second row of Fig ~\ref{ca-Line22}, from left to right, are the complete field sampling data, the irregular sampling data, the reconstruction results of SCL, the traditional self-supervised learning, and the SCRN method, respectively. It can be seen that the complete field sampling data contains a certain amount of random noise, which means that the reconstruction work of the irregular sampling data will be interfered with by the noise. Due to this interference, the results of the traditional self-supervised learning method contain a large number of reconstruction errors. Although these errors do not affect the judgment of large-scale geological targets, they cause great interference to the judgment of fine structures. In comparison, the results obtained using SCL and SCRN are more stable, with significantly fewer reconstruction errors than the traditional self-supervised learning method, and to a certain extent, the random noise contained in the data has been attenuated. The third row of Fig ~\ref{ca-Line22} shows a more complex area, and the reconstruction of seismic data for such complex areas is usually very challenging. Even so, the SCL we proposed can still have good performance, and the reconstruction results are very close to the complete field seismic data. Under such circumstances, the traditional self-supervised learning method fails to achieve satisfactory results. The resolution of the reflected wave signals in its reconstruction results is rather poor, and these errors will seriously affect the subsequent work of fine identification of underground structures. Although the overall reconstruction quality of the SCRN method is satisfactory, slight vertical artifacts can still be observed in the results. This is also evident in the residual maps, where SCRN exhibits more pronounced residuals. Fig.~\ref{trace_line22} shows the single-trace waveforms of the reconstructed data obtained by different methods. This can also be observed in the F–K spectra, where the energy of SCRN is more dispersed. It can be observed that, compared with the traditional self-supervised learning method and the SCRN method, the SCL results fit the original data more closely.

\begin{figure}
\centering
\noindent\includegraphics[width=\linewidth]{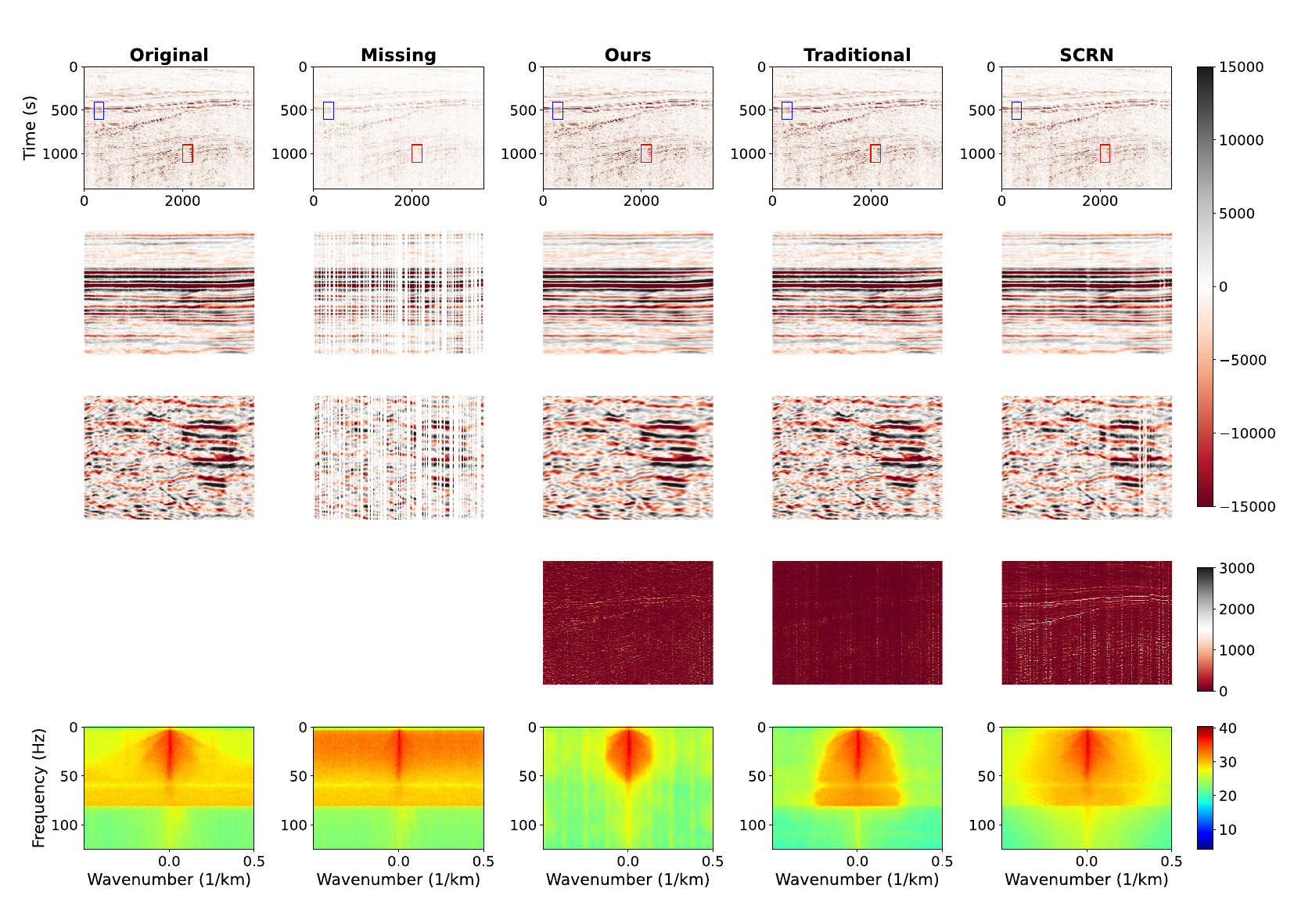}
\caption{The Line 33 survey line in the USGS National Petroleum Reserve–Alaska project. The first column represents the complete field collection results; the second column represents the irregular spatial sampling data used for reconstruction; the third column represents the reconstruction results of our proposed SCL; the fourth column represents the results of reconstruction using the traditional self-supervised learning method. The fifth column represents the results of reconstruction using the SCRN method. The first row presents the complete region, where the blue and red rectangles indicate two local regions enlarged in the second and third rows. The fourth row shows the residuals between the reconstructed results and the original data. The fifth row displays the corresponding F-K spectra for each case.}
\label{ca-Line33}
\end{figure}

\begin{figure}
\centering
\noindent\includegraphics[width=0.8\linewidth]{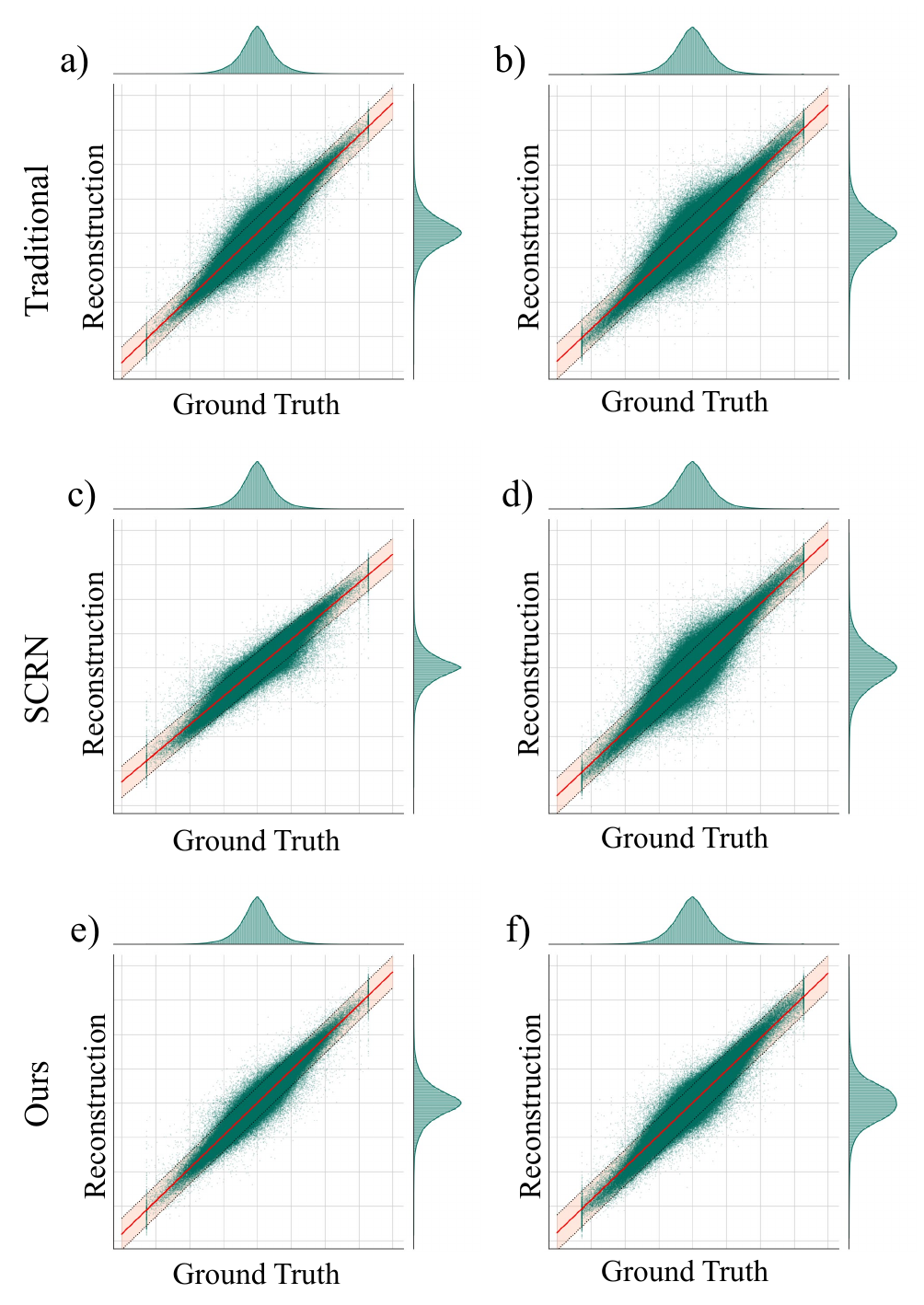}
\caption{Scatter plots of Line 22 and Line 33 data from the USGS Beaufort Sea–Arctic Alaska project. Panels a), c), and e) illustrate the scatter relationships between the reconstructed and the original data obtained by the traditional self-supervised learning method, SCRN, and SCL for line 22, respectively. Panels b), d), and f) show the corresponding results for line 33.}
\label{ca-scatter}
\end{figure}

\begin{figure}
\centering
\noindent\includegraphics[width=\linewidth]{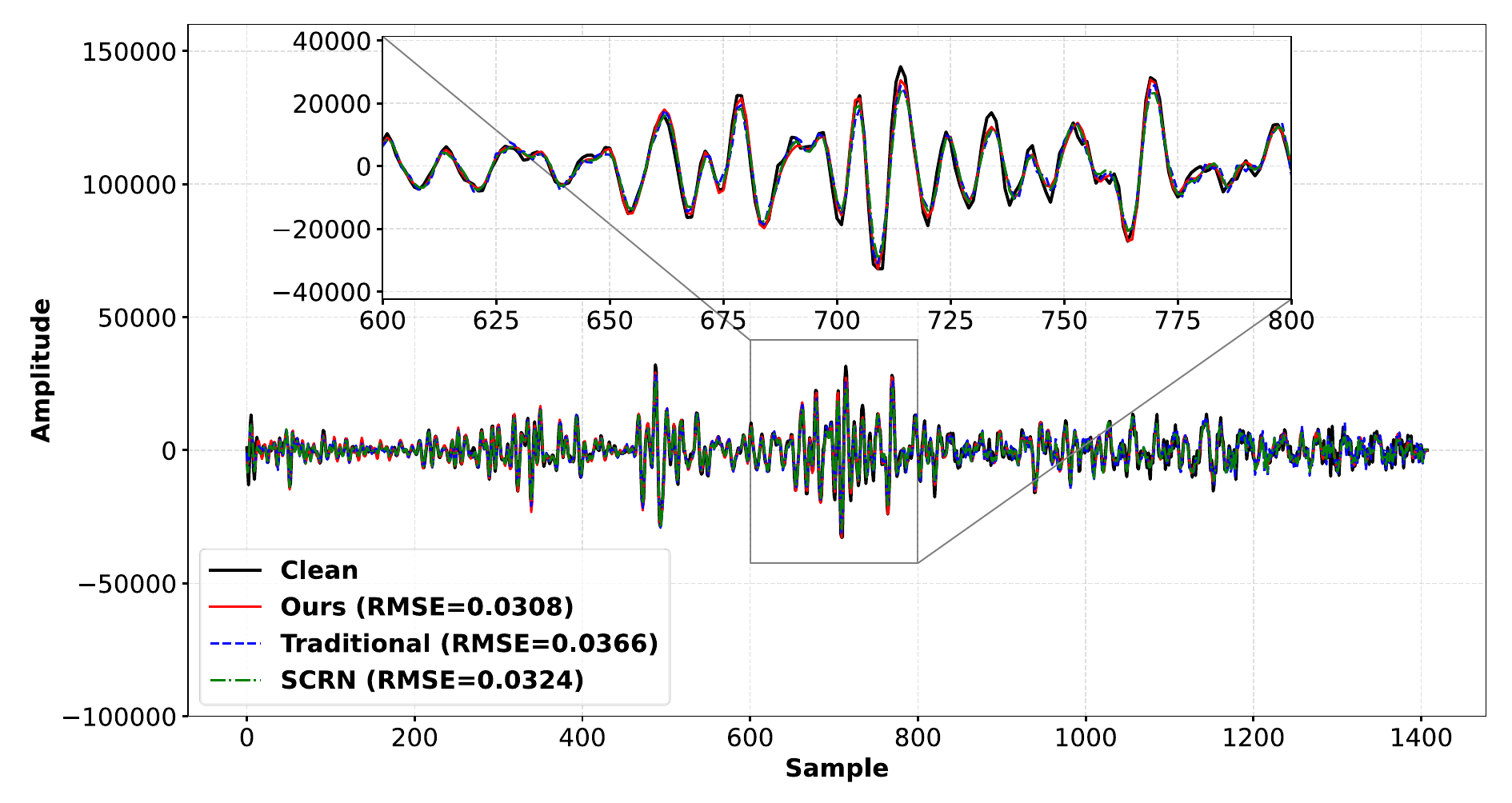}
\caption{Single-trace waveform (Trace 689) comparison of the reconstructed data by different methods for line 33.}
\label{trace_line33}
\end{figure}

In addition, the reconstruction results of survey line 33 are presented in Fig ~\ref{ca-Line33}. The experimental results of Line 33 show that although the traditional self-supervised learning method can reconstruct the strong reflection areas to a certain extent, the resolution of the reconstruction results in the surrounding weaker reflection areas is obviously insufficient. When reconstructing larger missing areas, it may lead to failure due to the lack of effective constraints, as seen in the red box area in Fig ~\ref{ca-Line33}. However, even when reconstructing such larger missing areas, SCL can demonstrate strong stability and basically reconstruct the shape of the reflected wave. However, when the SCRN method is applied to data with complex structures and severe continuous trace gaps, its reconstruction performance deteriorates, exhibiting pronounced vertical artifacts. Similarly, we have drawn the scatter plots of Survey Line 22 and Survey Line 33, as shown in Fig ~\ref{ca-scatter}. It can be clearly seen that the results of the traditional self-supervised learning and SCRN methods are more scattered overall, which means that the overall deviation is too large and the results are not stable enough. This greatly affects the credibility of the subsequent processing and interpretation results. In contrast, although there are still certain errors in the results of SCL to some extent, the deviation is generally within a stable range. It is more stable compared with the results of the traditional self-supervised learning and SCRN method. Likewise, Fig.~\ref{trace_line33} presents the single-trace waveform comparison between the original data and the reconstructed results obtained by different methods. All methods produce waveforms that closely match the original data, with the SCL method achieving the highest fidelity, followed by SCRN, and then the traditional self-supervised learning method.

In general, the experiments on two large-scale seismic exploration projects have fully demonstrated the advantage of SCL. It is more stable compared to traditional self-supervised learning and the SCRN method. The lightweight deep learning network enables it to efficiently process various large-scale seismic data, which is difficult for traditional computing methods and most deep learning methods to handle.

\section{Discussion}
\subsection{Comparison with Traditional Methods}

In this section, we selected four shorter survey lines to compare SCL with traditional methods, aiming to highlight the stability and effectiveness of SCL. We selected two of the most common traditional methods: the damped reduced-rank method \cite{54} \cite{57} and the fast dictionary learning method \cite{55} \cite{56}. The processing results of various methods are shown in Table ~\ref{compare}. It can be clearly seen that the SCL we proposed has achieved the best reconstruction effect on all four survey lines, and the effect is stable. The results of the traditional self-supervised learning method have great uncertainties. On some survey lines, the reconstruction performance is better than that of traditional methods, while on others, it lags. Both traditional methods can reconstruct seismic data to a certain extent, and their effects are relatively stable. However, their reconstruction performance needs to be further improved. In addition, this is not the main reason that restricts their application in large-scale seismic exploration data. In the next section, we will discuss in detail the issue of their computational efficiency. 

\begin{table}[!htbp]
\centering
\setlength{\tabcolsep}{2mm}
\renewcommand\arraystretch{1.2}
\begin{tabular}{cccccc}
\toprule[1.5pt]
\textbf{Survey Line} & & \textbf{SCL} & \textbf{\makecell{Traditional} } & \textbf{\makecell{Damped \\ Rank-Reduction}} & \textbf{\makecell{Fast Dictionary \\ Learning}} \\
\midrule
 & SNR & \textbf{10.4402} & 9.0981 & 8.1874 & 8.0301   \\
\textbf{WB-725m} & SSIM & \textbf{0.9206} & 0.9013 & 0.8669 & 0.8798 \\
 & $R^2$ & \textbf{0.9096} & 0.8769 & 0.8482 & 0.8426 \\
\midrule
 & SNR & \textbf{10.7898} & 7.2170 & 9.7962 & 7.7903 \\
\textbf{WB-901m} & SSIM & \textbf{0.9262} & 0.8509 & 0.8945 & 0.8863 \\
 & $R^2$ & \textbf{0.9166} & 0.8102 & 0.8952 & 0.8337 \\
\midrule
 & SNR & \textbf{10.4066} & 7.8538 & 9.5515 & 7.9724 \\
 \textbf{WB-905m} & SSIM & \textbf{0.9195} & 0.8750 & 0.9024 & 0.8870 \\
 & $R^2$ & \textbf{0.9089} & 0.8361 & 0.8891 & 0.8405 \\
\midrule
 & SNR & \textbf{12.1215} & 10.0342 & 8.2882 & 8.2797 \\
 \textbf{WB-919m} & SSIM & \textbf{0.9453} & 0.9155 & 0.8724 & 0.8933 \\
 & $R^2$ & \textbf{0.9386} & 0.9008 & 0.8517 & 0.8514 \\
\bottomrule[1.5pt]
\end{tabular}
\caption{The reconstruction results (with 50\% traces missing) of the four methods on the four shorter survey lines in the USGS Beaufort Sea–Arctic Alaska project. SCL values are highlighted in bold. (Bold values represent the best performance.)}
\label{compare}
\end{table}

\begin{figure}
\centering
\noindent\includegraphics[width=\textwidth]{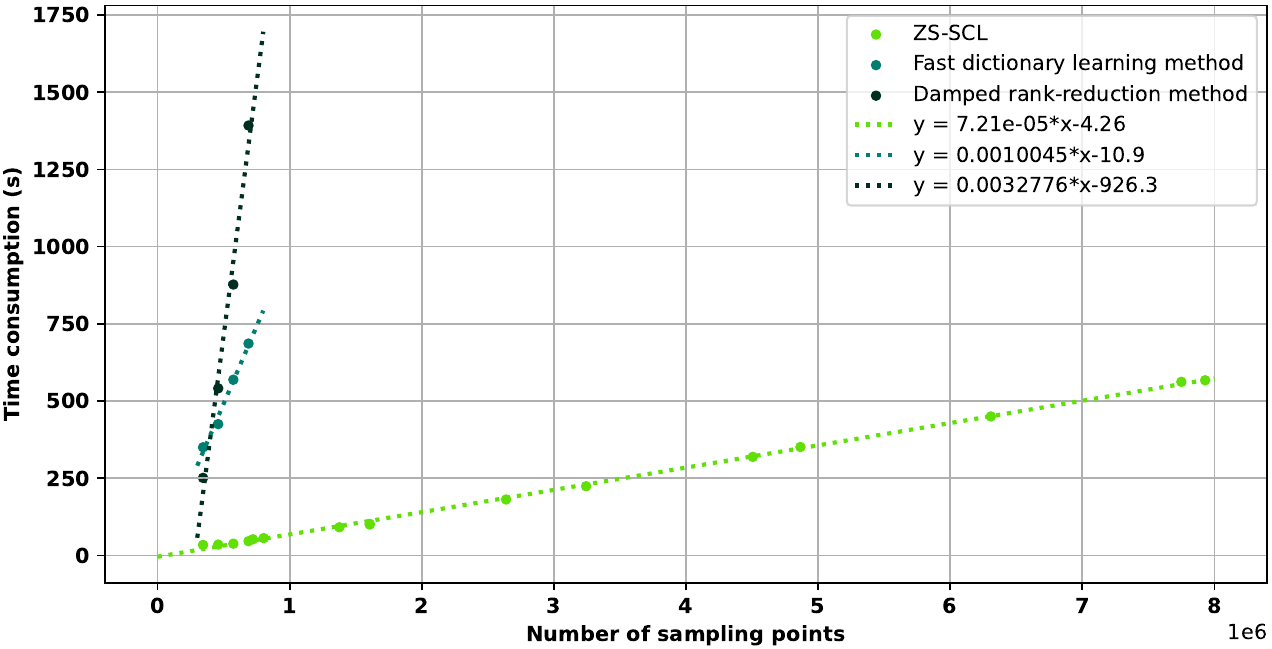}
\caption{The relationship between the processing time consumption of each survey line and the number of processing sampling points, where the number of sampling points is calculated through the grid size.}
\label{efficiency}
\end{figure}

\subsection{efficiency}
Processing efficiency is an indicator that cannot be ignored in data processing. For the task of reconstructing large-scale seismic data, the processing efficiency is usually extremely low. Various traditional computing methods have many parameters that are difficult to directly estimate and need to be set manually, which makes the whole processing procedure usually need to be carried out multiple times to determine the parameters. Besides, for large-scale data, some global optimal parameters used in traditional methods often fail to take local situations into account. One idea is to divide the data into multiple patches for separate processing; however, this will increase the burden of parameter selection. Many deep learning methods can achieve relatively high processing efficiency \cite{52} \cite{53}, but they are faced with the following several problems: Firstly, their processing effects largely depend on the training sets used for training. Secondly, they usually need to divide the large-scale seismic data into multiple patches for separate processing and then restore them. The way of dividing the patches will greatly affect their processing effects. 

The proposed SCL method effectively addresses two core challenges in seismic data reconstruction.
First, a self-supervised training strategy is employed, enabling the deep learning network to be trained without relying on any additional external data, while simultaneously improving the reconstruction resolution compared with traditional self-supervised learning methods. In addition, since a lightweight network architecture is adopted, the proposed method can directly process large-scale seismic data without dividing it into small patches, thereby avoiding the potential artifacts and inconsistencies caused by patch-based processing.

The experiments were conducted on an NVIDIA RTX~4060 server equipped with 8 GB of memory. As shown in Fig.~\ref{efficiency}, the processing time for a single survey line is significantly reduced. Even when processing a seismic line containing approximately $8\times10^8$ sampling points (equivalent to 4000 receivers recording 8~seconds of data at a 4~ms sampling interval), the complete training and reconstruction process of SCL takes less than 580~seconds, whereas traditional methods require several hours. Furthermore, the time consumption analysis in Fig.~\ref{efficiency} reveals an overall linear growth trend with respect to data size, indicating that the SCL method maintains high computational efficiency and does not exhibit exponential growth in computation time as the dataset size increases. We also plotted the processing time of the two traditional calculation methods on the four shorter survey lines, and we used the Intel Core i5 14400F CPU for the computation. It can be seen that, compared with SCL, the efficiency of traditional computing methods is dozens of times lower. Although it can be applied to the reconstruction of some small-scale seismic exploration data, it is difficult to directly apply it to large-scale seismic exploration data. 

\subsection{Ablation of loss function}
\begin{figure}
\centering
\noindent\includegraphics[width=0.5\textwidth]{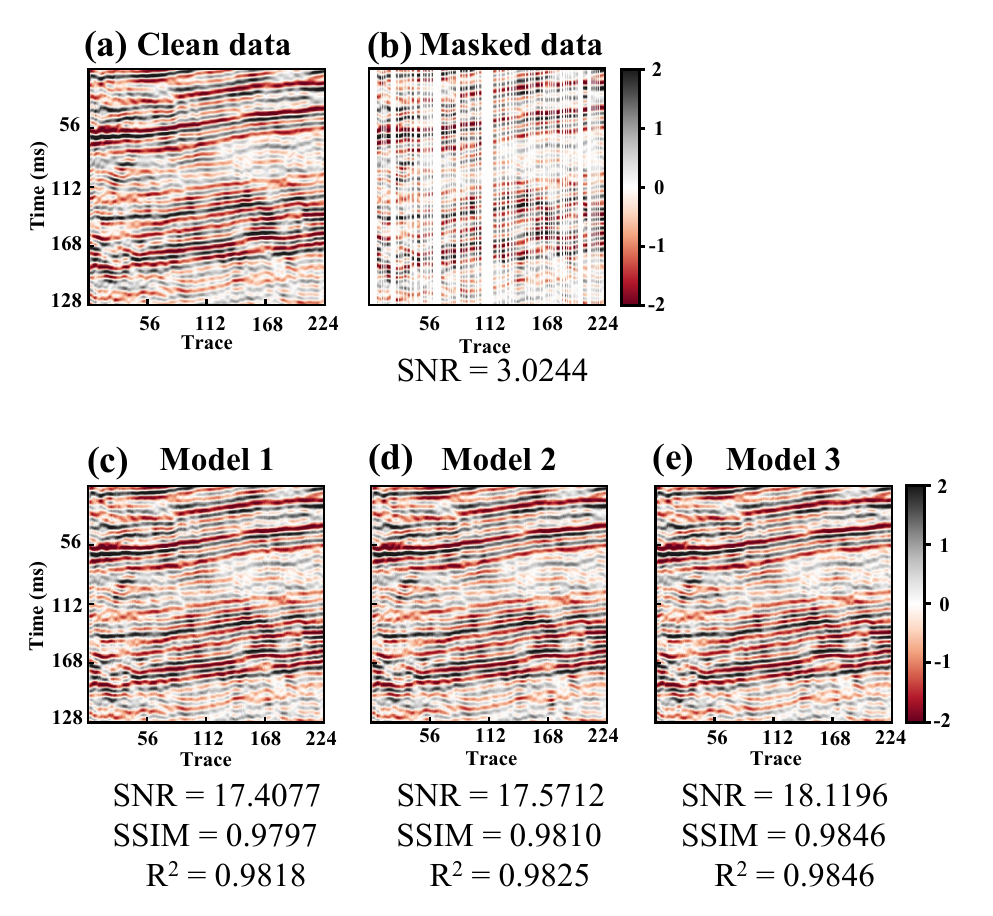}
\caption{Ablation study of the loss function: comparison among Model 1, Model 2, and Model 3.}
\label{ablation}
\end{figure}

To intuitively demonstrate the advantages of our strategy, we conducted an ablation study on our loss function (Equation 9).
Equation 9 can be written as
\begin{equation}
\theta = \arg \min_\theta \Big( \underbrace{\|d-N(d|\theta)*R\|_2^2}_{\text{Term 1}} + 
\underbrace{\|d-N(N(d|\theta)*R^{'})|\theta)*R\|_2^2}_{\text{Term 2}} + 
\underbrace{\|N(d|\theta)-N(N(d|\theta)*R^{'}|\theta)\|_2^2}_{\text{Term 3}} \Big).
\end{equation}
In the ablation study, Model 1 corresponds to the first term, representing the traditional self-supervised learning method. Model 2 corresponds to the sum of the first and second terms, and Model 3 includes all three terms of the complete loss function, representing the SCL method. As shown in Figure~\ref{ablation}, we present the reconstruction results of the three models and evaluate them using three metrics. Compared to Model 1 and Model 2, Model 3 achieves higher values on all three metrics, demonstrating the effectiveness of our proposed method.

\section{Conclusion}

In this study, we propose a deep learning method for reconstructing seismic data with irregular spatial sampling. The method employs a self-supervised learning strategy, enabling a lightweight deep learning network to be trained without requiring additional datasets. Moreover, we introduce a self-consistency learning strategy that effectively improves the network's modeling accuracy and stabilizes the reconstruction process. The use of a lightweight network itself is also an advantage, as it significantly enhances computational efficiency when handling large-scale data and eliminates the need to divide the data into patches. We validate the proposed method on two large-scale seismic field datasets. Compared with traditional self-supervised deep learning strategies and traditional methods, our approach demonstrates superior stability, reconstruction performance, and computational efficiency, making it highly valuable for large-scale seismic exploration projects.

\section{Acknowledgments}

The code of SCL, research lines, and the results of different comparative methods have been uploaded to a GitHub repository \url{https://github.com/lexiaoheng/Zero-Shot-Self-Consistency-Learning-for-Seismic-Reconstruction}.

\section*{conflict of interest}
There is no conflict of interest between authors.

\printendnotes

% Submissions are not required to reflect the precise reference formatting of the journal (use of italics, bold etc.), however it is important that all key elements of each reference are included.
\bibliography{sample}

\end{document}